\documentclass[twocolumn,english,conference]{IEEEtran}
\usepackage[T1]{fontenc}
\usepackage{adjustbox}
\usepackage{graphicx}
\usepackage{calc}
\usepackage{mdwlist}
\usepackage{amsthm}
\usepackage{amstext}
\usepackage[numbers,square,comma]{natbib}
\usepackage{graphicx}
\usepackage{caption}
\usepackage{subcaption}
\usepackage {dsfont}
\usepackage{algorithm} 
\usepackage{algpseudocode}
\usepackage[utf8]{inputenc}
\renewcommand{\algorithmicrequire}{\textbf{Inputs:}}
\renewcommand{\algorithmicensure}{\textbf{Output:}}
\usepackage{color}
\usepackage{enumerate}

\makeatletter

\DeclareRobustCommand*{\lyxarrow}{%
\@ifstar
{\leavevmode\,$\triangleleft$\,\allowbreak}
{\leavevmode\,$\triangleright$\,\allowbreak}}

\theoremstyle{plain}

\theoremstyle{plain}

\usepackage[caption=false,font=footnotesize]{subfig}
\usepackage{bbold}
\usepackage{dsfont}

\usepackage[footnote,nomargin]{fixme}

\newcommand{\fnsv}[1]{\fxnote{SV: #1}}

\makeatother

\providecommand{\lemmaname}{Lemma}
\providecommand{\theoremname}{Theorem}

\begin{document}

\title{Inter-User Interference Coordination in Full-Duplex Systems Based on Geographical Context Information}
\author{\IEEEauthorblockN{Melissa Duarte, Afef Feki, and Stefan Valentin}
\IEEEauthorblockA{Mathematical and Algorithmic Sciences Lab,\\
FRC, Huawei Technologies, Paris, France,\\
{\{melissa.duarte, afef.feki, stefan.valentin\}@huawei.com}}
}

\maketitle

\begin{abstract}
	We propose a coordination scheme to minimize the interference between users in a cellular network with full-duplex base stations and half-duplex user devices. Our scheme exploits signal attenuation from obstacles between the users by (i) extracting spatially isolated regions from a radio map and (ii) assigning simultaneous co-channel uplink and downlink transmissions to users in these regions such that inter-user interference is minimized. While adding low computational complexity and insignificant signaling overhead to existing deployments, evaluating our solution with real coverage data shows impressive gains compared to conventional half-duplex and full-duplex operation.

\end{abstract}


\section{Introduction}\label{sec:intro}
Base stations can considerably increase spectral efficiency by enabling simultaneous uplink and downlink transmission within the entire frequency band \cite{SAB14}. Such in-band full-duplex operation avoids the orthogonal use of channel resources and can, thus, increase capacity. A main problem of such operation, however, is the interference between the simultaneous co-channel uplink and downlink transmissions of neighboring users. 
In this paper, we introduce an interference coordination scheme that minimizes such inter-user interference by exploiting the signal attenuation due to obstacles between the users. This attenuation is extracted from geographical context information, provided by radio maps \cite{MOM,REM,7152980} and user positions.

\subsection{Idea and Contributions}
To minimize the interference between full duplex users, the proposed coordination scheme exploits electromagnetic isolation. Even neighboring users may be separated by buildings or other obstacles, which leads to a high attenuation of the direct path between them. Our coordination scheme detects the spatial isolation, provided by such attenuators, and assigns uplink and downlink transmission such that inter-user interference is minimized. 

To do so, our scheme (i) extracts spatially isolated areas from a radio map, (ii) estimates the attenuation between these areas, and (iii) solves an assignment problem to minimize the interference between users located in these areas. Our simulation results verify the effectiveness of our solution and show high gains in spectral efficiency, compared to baseline half-duplex operation and full-duplex with random coordination. For the operating network, these gains come only at the cost of obtaining the users' positions which is either already supported or introduces low signaling overhead \cite{1018015}. Since the radio maps can be processed offline, the only significant online operation corresponds to an assignment problem.

\subsection{Related Work}
There is a plethora of methods to reduce interference in cellular networks. Almost Blank Subframe (ABS)~\cite{PED12} use orthogonal resource allocation patterns to reduce the interference at the cell edge. Although practical and widely adopted in cellular networks, these schemes require additional signaling to coordinate the ABS patterns and, above all, reduce spectral efficiency by introducing orthogonal transmission in time. An alternative like Fractional Frequency Reuse (FFR)~\cite{NOV10} reduces spectral efficiency of the system because of the orthogonal frequency allocation for neighboring users, which increases the distance between co-channel users. This prevents the use of full-duplex in the center of the cell. Power control~\cite{KUR14} can reduce the interference from uplink or downlink users but decreases the received power and, thus, spectral efficiency. 


Although Interference Alignment (IA) \cite{SUN14,SAH13-2} does not reduce the spectral efficiency, it requires accurate Channel State Information (CSI) between communicating and interfering users. Such CSI comes at the cost of high signaling overhead. Since interference alignment operates on a number of time slots that depends on the CSI, it also adds a non-deterministic transmission delay. Further solutions require multiple antennas at the user equipment (UE) \cite{SUN14}. As ABS and FFR, such approaches use the channel's degrees of freedom for interference compensation instead of data transmission and, thus, reduce spectral efficiency. Other solutions make use of additional bandwidth to establish a side channel between interfering users \cite{BAI13}. With limited bandwidth this means, again, a reduction of spectral efficiency.

In summary, the above alternatives either reduce spectral efficiency of the wireless link by introducing orthogonality or requiring further resources or come at the cost of high signaling overhead and significant delay. During operation, our interference-coordination scheme adds only insignificant overhead to existing cellular networks in order to obtain the users' positions. Moreover, our scheme does not decrease spectral efficiency, adds no transmission delay and can be applied in the cell center.

\subsection{Structure}
Section II of this paper presents the proposed interference-coordination scheme. After introducing system model and parametrization, Section III studies our scheme's performance in a simple scenario based on real and simulated propagation measurements. Finally, Section IV concludes the paper.

\begin{figure*}[t]
    \centering
    \begin{subfigure}[t]{0.3\textwidth}
        \centering
        \includegraphics[scale=1.5]{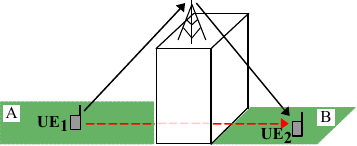}
        \caption{Macro-cell scenario. Base station on top of a building and users at opposite sides of a building.}
        \label{fig:ObstructionMacro}
    \end{subfigure}\qquad 
    \begin{subfigure}[t]{0.3\textwidth}
        \centering
        \includegraphics[scale=1.2]{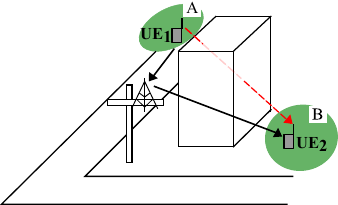}
        \caption{Small cell scenario. Base station on a lamp post and users at opposite sides of a building.}
        \label{fig:ObstructionSmall}
    \end{subfigure}\qquad
    \begin{subfigure}[t]{0.3\textwidth}
        \centering
        \includegraphics[scale=1.2]{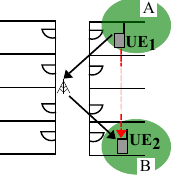}
        \caption{Indoor scenario. Access point in the corridor and users separated by walls.}
        \label{fig:ObstructionIndoor}
    \end{subfigure}
    \caption{Example scenarios where spatially separated areas can be exploited for interference coordination. The interference created by an uplink user UE$_1$ in region $A$ to a downlink user UE$_2$ in region $B$ is attenuated by obstructions like buildings and walls.}
    \label{fig:Obstructions}
\end{figure*}

\section{Inter-User Interference Coordination Based on Geographical Context}
In this section, we propose a framework for inter-user interference coordination tailored for full-duplex operation and based on geographical context information. First, we introduce a metric to characterize the attenuation between geographical regions. We name this metric the mitigation factor $\alpha$. Second, we present a method to identify spatially isolated regions, to compute $\alpha$ between them, and to store the result in a database. Third, we explain how to use this database to identify spatially isolated users and to perform interference coordination for a full-duplex base station.

\begin{figure*}[t]
    \centering
    \begin{subfigure}[t]{0.22\textwidth}
        \centering
        \includegraphics[height=1.5in]{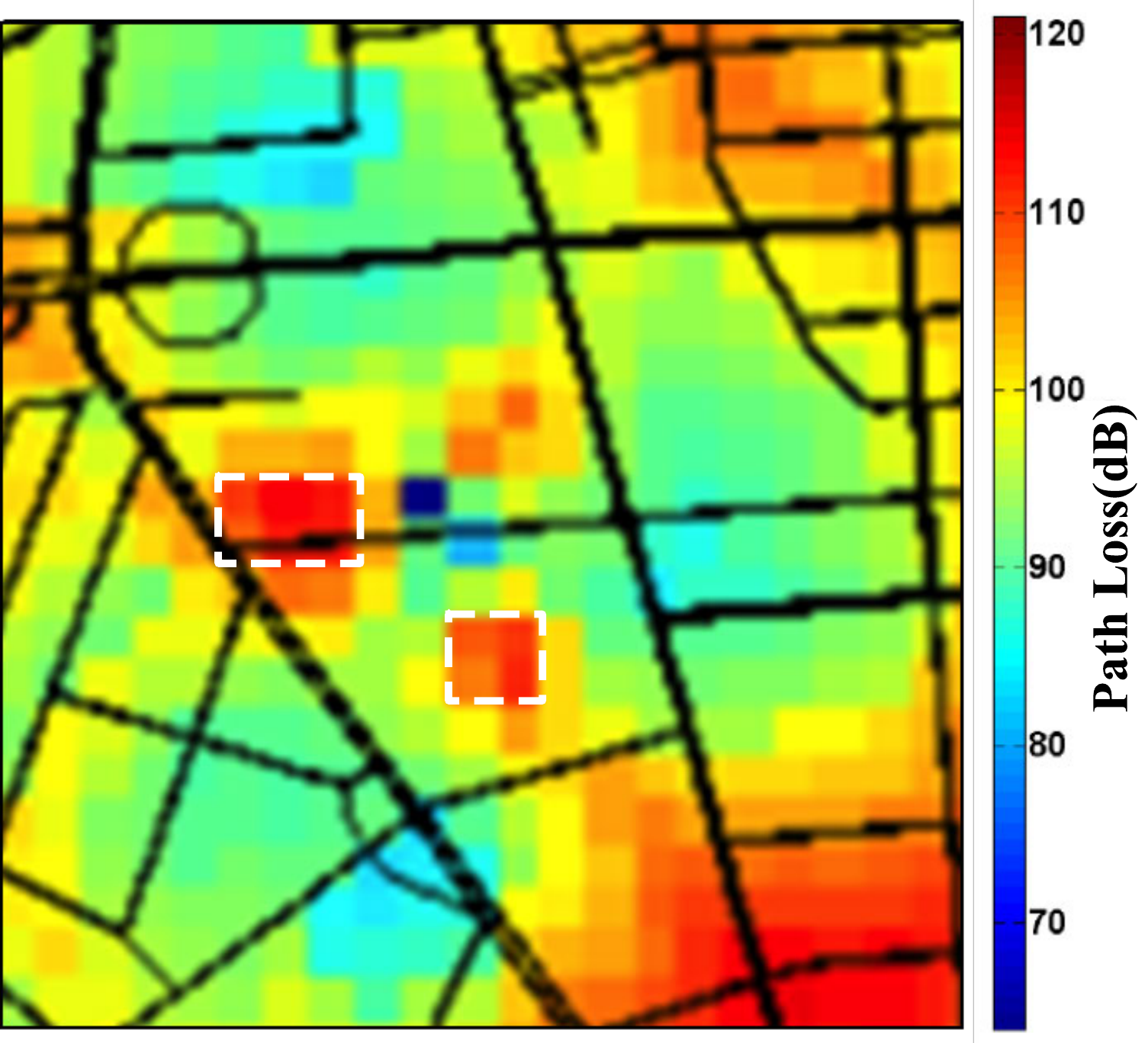}
        \caption{Two areas of high path loss (delimited by dashed white boxes) are identified on a radio map.}
        \label{fig:RadioMapObstructions}
    \end{subfigure}\qquad 
    \begin{subfigure}[t]{0.22\textwidth}
        \centering
        \includegraphics[height=1.5in]{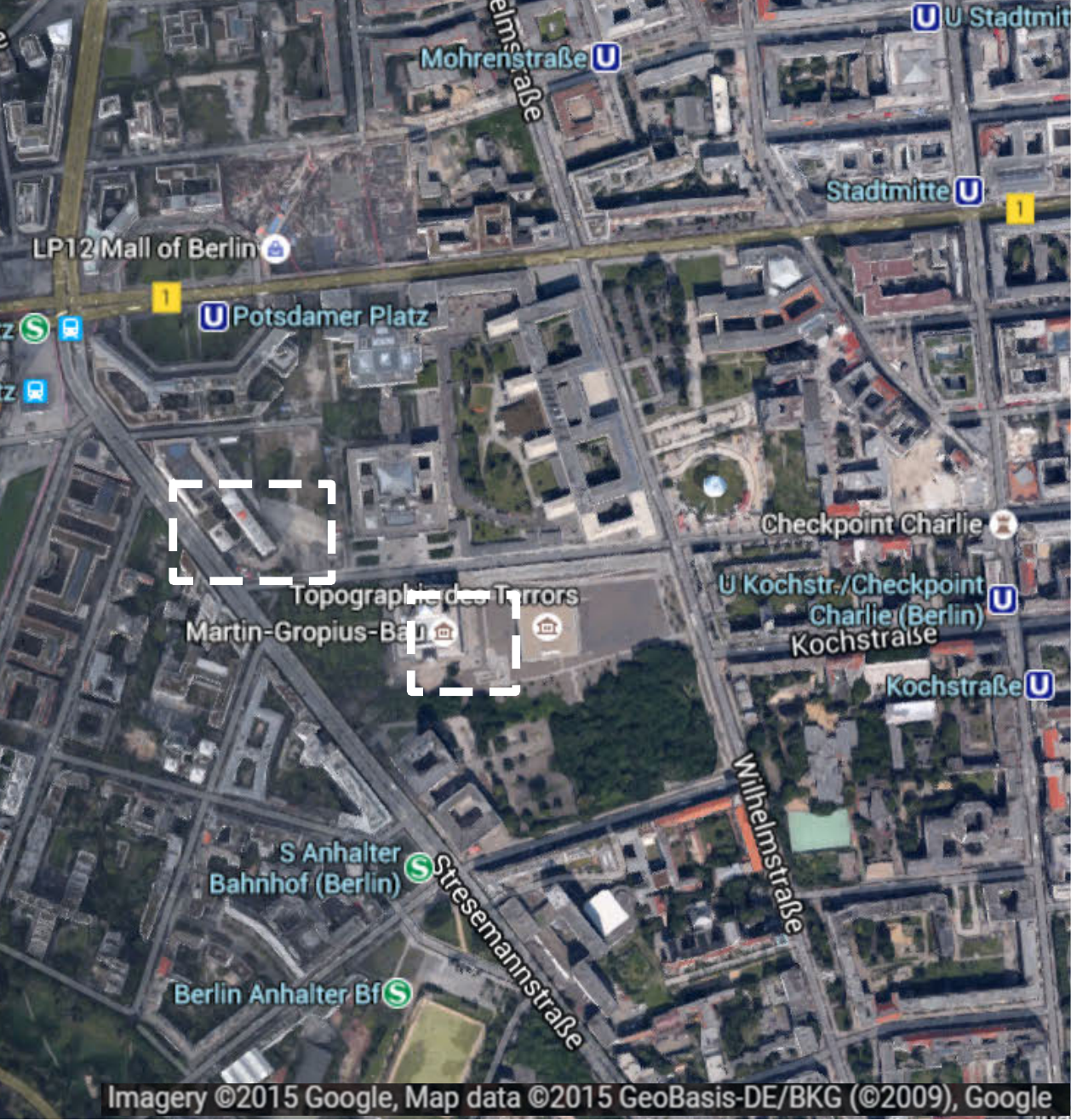}
        \caption{We can verify by a street map that the two areas in Fig. \ref{fig:RadioMapObstructions} correspond to buildings.}
        \label{fig:StreetMapObstructions}
    \end{subfigure}\qquad
    \begin{subfigure}[t]{0.22\textwidth}
        \centering
        \includegraphics[height=1.55in]{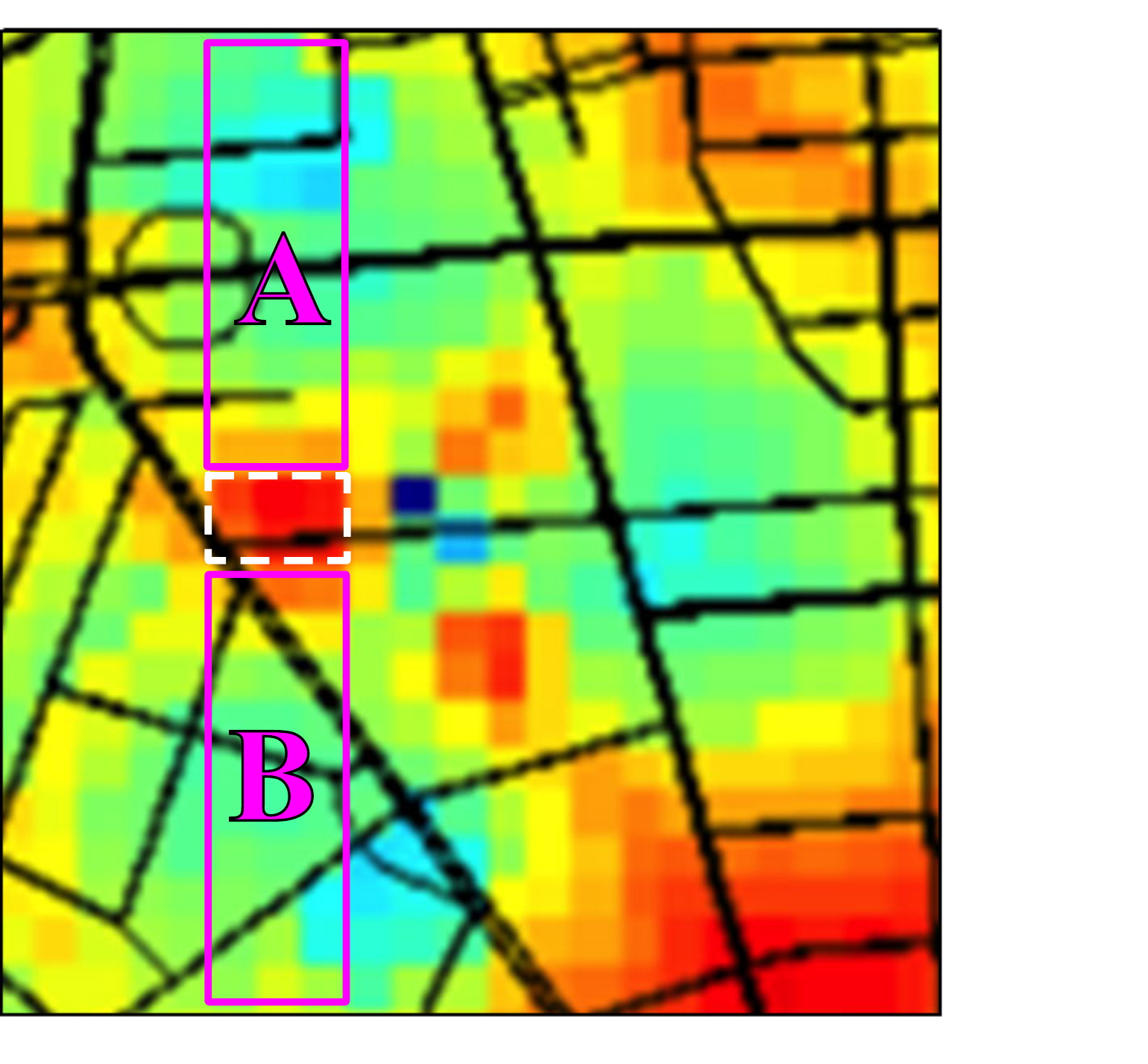}
        \caption{The areas from Fig. \ref{fig:RadioMapObstructions} are used to define the adjacent spatially isolated regions $A$ and $B$.}
        \label{fig:RadioMapIsolatedRegions}
    \end{subfigure}\qquad
    \begin{subfigure}[t]{0.22\textwidth}
        \centering
        \includegraphics[height=1.55in]{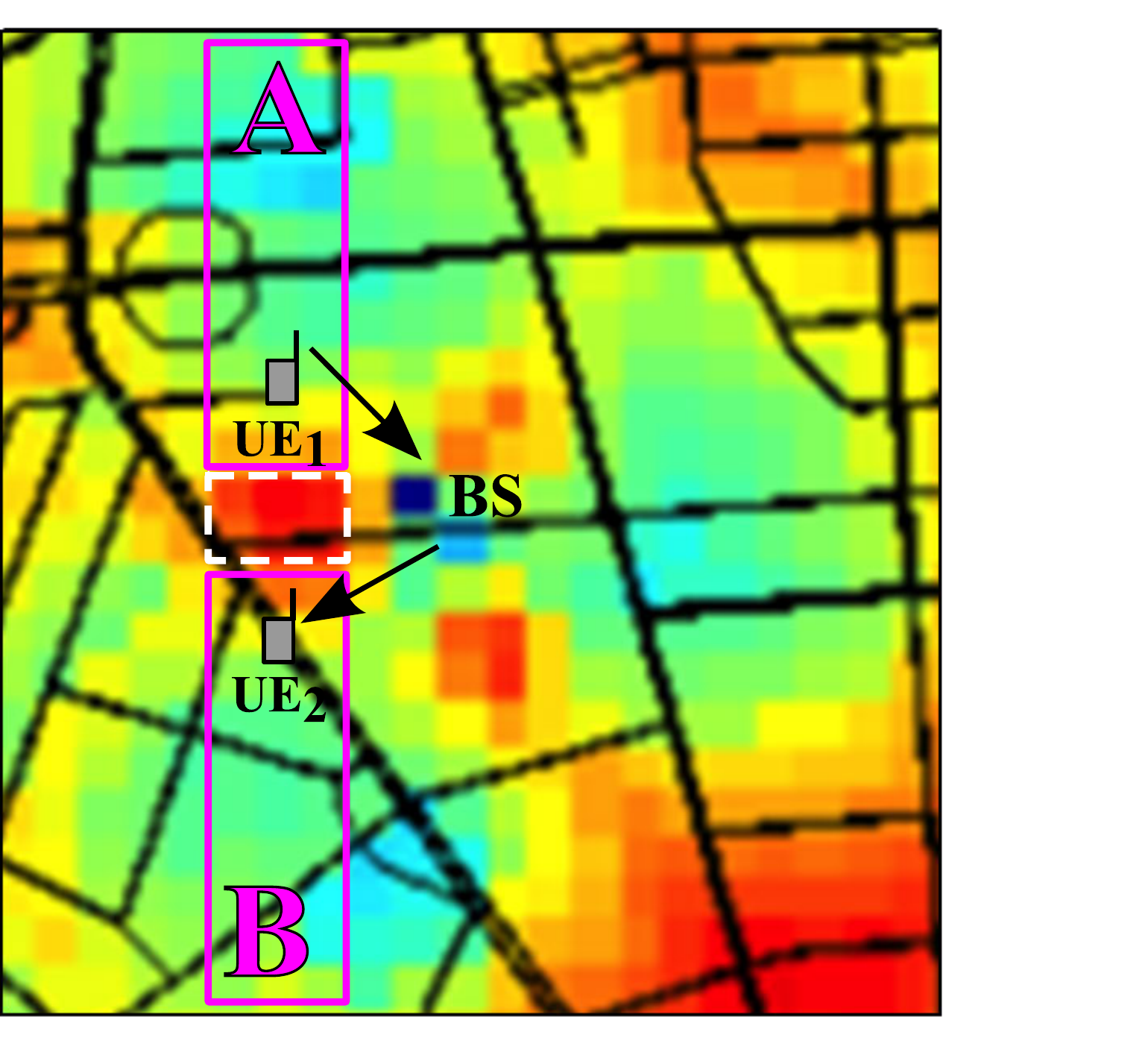}
        \caption{Now uplink and downlink transmissions can be assigned to users within $A$ and $B$ such that inter-user interference is minimized.}
        \label{fig:RadioMapIsolatedUsers}
    \end{subfigure}
    \caption{Using a radio map from \cite{MOM} for the identification of spatially isolated regions that enable the proposed inter-user interference mitigation. Note that the step in Fig. \ref{fig:StreetMapObstructions} is not part of the procedure and only added for the sake of explanation.}
    \label{fig:RadioMap}
\end{figure*}

\subsection{Mitigation Factor}

Inter-user interference mitigation can be achieved by exploiting the location of obstructions that attenuate the wireless signal. Examples of such obstructions and their effect on inter-user interference are shown in Fig.~\ref{fig:Obstructions}. In Fig.~\ref{fig:ObstructionMacro} and Fig.~\ref{fig:ObstructionSmall}, the inter-user interference that user UE$_1$ generates to user UE$_2$ is attenuated by a building that is located between the two users. In Fig.~\ref{fig:ObstructionIndoor}, the inter-user interference between UE$_1$ and UE$_2$ is attenuated by the walls between the users.

Let us define the regions $A$ and $B$ that are separated by such an attenuator, as illustrated in Fig.~\ref{fig:Obstructions}. We characterize the interference attenuation between different regions as the mitigation factor $\alpha$. This factor indicates the minimum level of signal attenuation between regions $A$ and $B$. Formally defined: For any user UE$_1 \in A$ and any user UE$_2 \in B$, the inter-user interference mitigation between UE$_1$ and UE$_2$ is greater than $\alpha$ dB. Hence, for a given pair of regions $A$ and $B$, the mitigation factor $\alpha$ between these two regions indicates that the attenuation of a signal transmitted from $A$ and received at $B$ is at least $\alpha$ dB.


\subsection{Creating a Database for Spatial Isolation}
\label{subsec:DatabaseCreation}
We use spatially isolated regions to assign co-channel uplink and downlink users. Let us now describe the extraction of these regions and the calculation of the mitigation factor $\alpha$ in detail. The result of this procedure, which is performed offline on a potentially large radio map, is then saved in a database. Then, this database is used for the interference coordination scheme, operating in real time.

\textit{STEP 1. Maps are used to identify obstructions that attenuate the propagation of a wireless signal.} A high path loss in a radio map can indicate the presence of an obstruction, for example, a building, as shown in Fig.~\ref{fig:RadioMapObstructions} and Fig~\ref{fig:StreetMapObstructions}. While obstructions can also by identified from street maps and building plans, in this paper only radio maps, on which areas of strong attenuation can be identified, are required. Note that constructing radio maps is out of the scope of this paper, but many studies such as \cite{MOM,REM,7152980} deal with this topic.

\textit{STEP 2. Identify spatially isolated regions around an obstruction.}
Spatially isolated regions are geographical areas with high signal attenuation between them. Fig.~\ref{fig:RadioMapIsolatedRegions} illustrates the process of identifying the spatially isolated regions $A$ and $B$. Once the path loss within a region exceeds a threshold, the region can be specified as an area within $d$ meters away from the obstruction boundary. The segmentation of a radio map into such regions can be performed by image processing techniques such as~\cite{ARB11}.

\textit{STEP 3. Mitigation factor $\alpha$ between regions is computed.}
To compute the signal attenuation $\mu(p_a,p_b)$ between any position $p_a$ in $A$ and any position $p_b$ in $B$, one of the following methods can be used: (i) a path loss model (ii) using ray tracing (iii) measures of area as given in a radio map (iv) measures at dedicated positions, e.g., from drive tests. Let $\mathcal{M}$ be the set containing all the computed values of $\mu(p_a,p_b)$ for $A$ and $B$. The mitigation factor $\alpha$ between $A$ and $B$ is equal to $\min_{\mathcal{M}}\mu(p_a,p_b)$. Notice that the larger the set $\mathcal{M}$, the higher resolution one will have for the attenuation between regions $A$ and $B$. If there is a pair of points $p_a$ and $p_b$ for which $\mu(p_a,p_b)$ is less than a predefined threshold value, then one can go back to STEP~2 and redefine the region boundaries.

\textit{STEP 4. Storing spatial isolated regions in a database.}
If the mitigation factor $\alpha$ computed in STEP~3 satisfies a predefined threshold condition then the value of $\alpha$ and the corresponding region localization are stored in a database. There may be multiple pairs of spatially isolated regions around an obstruction and more than one obstruction in a given area. To organize this information, we use a  database in the form of pairs of regions and their corresponding mitigation factor. The $k$-th entry in the database corresponds to the $k$-th pair of spatially isolated regions and the entry is a triplet that has (1) the mitigation factor $\alpha_k$ between the regions $A_k$ and $B_k$ (2) the positions of the boundaries that delimit region $A_k$ and (3) the positions of the boundaries that delimit region $B_k$. The database can be stored either at the base station or a central controller in the core network.

\subsection {Interference coordination scheme: Problem statement}
Let us now describe how the database with the mitigation factors is used by the network to coordinate the uplink and downlink transmission such that inter-user interference is minimized. 

First, the Base Station (BS) has to obtain the positions of the active users with outstanding scheduling grants. These positions can often be obtained from an existing localization service in the cellular network \cite{1018015}. Based on their positions, the users attached to a base station are separated into two sets: the set that belongs to the spatial isolated regions and the set of users outside these regions. Transmissions of users in the second set are allocated by the conventional BS scheduler. However, users in the first set are handled first by our interference coordination scheme. 

The idea of this scheme is illustrated in Fig. \ref{fig:RadioMapIsolatedUsers} and we can formulate this allocation of uplink and downlink transmissions as the following assignment problem between the set of transmissions (including both uplink and downlink) and the set of available resources (frequency resources per transmission time interval in our case).
\begin{eqnarray}
min & \sum_{u=1}^{|U|}{\sum_{f=1}^{|F|}{c_{u,f} x_{u,f}}} \label{mincond}\\
s.t. & \sum_{u=1}^{|U|}{x_{u,f}}=2 & \forall f \label{FDcond1}\\
		 & \sum_{u=1,\textrm{u odd}}^{|U|}{x_{u,f}}=1 & \forall f \label{FDcond2}\\
		 & \sum_{u=1,\textrm{u even}}^{|U|}{x_{u,f}}=1 & \forall f \label{FDcond3}\\
		 & \sum_{f=1}^{|F|} {x_{u,f}}=1 & \forall u \label{UEcond}\\
		 & x_{u,f} \in \left\{0,1\right\} & \forall u,f \label{Valcond}
\end{eqnarray}
Therein, $c_{u,f}$ corresponds to the cost measure for a user $u$ using frequency resource $f$, $|U|$ is the total number of users to schedule and $|F|$ is the total available frequencies. $x_{u,f}$ is the assignment variable that equals to $1$ if user $u$ is assigned frequency $f$ and $0$ in the contrary. 

The constraint (\ref{UEcond}) requires that a user, either in uplink or downlink, is assigned only one frequency, while the constraints (\ref{FDcond1})-(\ref{FDcond3}) state that only two users can be assigned to one frequency. This represents full-duplex operation and uses odd numbers of $u$ to refer to users in downlink operation while an even $u$ refers to the users in uplink. This notation avoids the use of a third dimension and, thus, allows to apply bipartite matching algorithms. A common method to find optimal bipartite matches is the Hungarian algorithm, which has the computational complexity of $\mathcal{O}(|U|^3)$ \cite{Book09}. Note that this is typically much faster than solving (\ref{mincond})-(\ref{UEcond}) by Binary Integer Programming, which is NP-hard.

In the objective formulation (\ref{mincond}), the cost measure
\begin{equation}
c_{u,f}= \frac{1}{r\left(\frac{\mbox{S}_u}{N_u+I_{v\rightarrow u, f}}\right)}
\label{cost}
\end{equation}
is defined as the inverse of the rate experienced by the user $u$ when communicating on frequency resource $f$. The rate $r$ is a function of the Signal-to-Interference-plus-Noise Ratio (SINR), which is composed of the received signal power $\mbox{S}_u$ for user $u$, the noise power $N_u$ that affects user $u$'s signal, and the interference power $I_{v\rightarrow u, f}$ at user $u$ due to the transmission of user $v$ on frequency $f$. Exact real-time knowledge of $I_{v\rightarrow u, f}$ is prohibitively complex. Thus, we use the knowledge of the regions and the users' positions to compute an approximation of  $I_{v\rightarrow u, f}$ as:
\begin{eqnarray}
I_{v\rightarrow u, f} & = &  P_{v} - \max\left(\left(\mathds{1}_{FDreg} \times \alpha\right), p_{v,u}\right),
\end{eqnarray}
where $P_{v}$ is user $v$'s transmission power in dBm, $p_{v,u}$ is the path loss between user $v$ and user $u$ and $\mathds{1}_{FDreg}$ is an indicator function that equals to $1$ if the users $v$ and $u$ are spatially isolated users and equals to $0$ otherwise. Since the position of the users is known, the path loss $p_{v,u}$ between two users $v$ and $u$ can be approximated via mathematical models \cite{3GPP1,3GPP2}. 

Notice that constraints (\ref{FDcond1})-(\ref{FDcond3}) implicitly assume that (i) the number of DL users to schedule is equal to the number of UL users to schedule and (ii) the number of available frequency resources $|F|$ is at least $|U|/2$.  These assumptions allow for a simplified problem statement but can be modified in future work to consider a more general case.

\subsection {Interference coordination scheme: Assignment heuristics}
\label{subsec:heurstics}
For a large amount of users, the Hungarian algorithm, although optimal, may not be computational efficient \cite{Book09}. Hence, we propose two  heuristics to approximate the solution of the assignment problem described in the previous section. These two heuristics, which we label as \textit{FDregrand} and \textit{FDregHDelse}, are explained below.

\subsubsection{FDregrand}
The BS uses FD in all available frequency resources. The BS first assigns frequencies to users that are in the identified pairs of spatially isolated regions. Then, users outside the regions are assigned frequencies randomly. The pseudocode for \textit{FDregrand} is shown in Algorithm 1.

\subsubsection{FDregHDelse}
The BS uses FD only in a subset of available frequency resources. The pairs of isolated regions are used to decide which users to serve in FD UL-DL. Whereas, users outside these regions are served in HD UL-DL with a random frequency assignment. The pseudocode for \textit{FDregHDelse} procedure is shown in Algorithm 1.

\begin{algorithm}
\caption{Proposed full duplex operation}
\label{FDalgo}
\algorithmicrequire{ Region pairs $\left(A_k, B_k\right)$, $k=1,2, ... K.$\\
$~~~~~~~~~~$ Set of users $U$.\\
$~~~~~~~~~~$ Set of available frequencies resources $F$.
} \\
\algorithmicensure{ $x_{u,f},\  \forall u \in U$ and $\forall f \in F$.} \\
\textbf{Initialize:} $x_{u,f}=0\   \forall u,\  \forall f$.\\
$~~~~~~~~~~~~~$ Initialize frequency resource counter: $f=0$. \\
\textbf{Step 1: Scheduling of users inside regions.}\\
\textbf{for} $1 \leq k \leq K$
\begin{enumerate}[$\hspace{10pt}$]
\item[(a)] Find DL users in region $A_k$: $\textrm{DL}_{A_k} = \left\{u:   u_{odd} \in A_k \right\}$. \\
Find UL users in region $B_k$: $\mathrm{UL}_{B_k} = \left\{u:   u_{even} \in B_k \right\} $.\\
	Update frequency resource initial index: $f_0=f$.\\
\textbf{for} $ 1 \leq l \leq \min(|\textrm{DL}_{A_k}|,|\textrm{UL}_{B_k}|)$\\
- Update frequency resource counter:$f = f_0+l$.\\
- Assign full-duplex mode: assign frequency resource $f$ to a DL user and an UL user, $x_{\textrm{DL}_{A_k}(l),f} = 1$, $x_{\textrm{UL}_{B_k}(l),f} = 1$.\\
\textbf{endfor}
\item [(b)] Find UL users in region $A_k$: $\textrm{UL}_{A_k} = \left\{u:   u_{even} \in A_k \right\}$. \\
Find DL users in region $B_k$: $\mathrm{DL}_{B_k} = \left\{u:   u_{odd} \in B_k \right\} $.\\
	Update frequency resource initial index: $f_0=f$.\\
\textbf{for} $ 1 \leq l \leq \min(|\textrm{UL}_{A_k}|,|\textrm{DL}_{B_k}|)$ \\
- Update frequency resource counter:$f = f_0+l$.\\
- Assign full-duplex mode: assign frequency resource $f$ to a DL user and an UL user, $x_{\textrm{UL}_{A_k}(l),f} = 1$, $x_{\textrm{DL}_{B_k}(l),f} = 1$.
\textbf{endfor} 
\end{enumerate}
\textbf{endfor} \\
\textbf{Step 2: Scheduling of users outside regions and users inside regions that were not scheduled in Step 1.}
\begin{enumerate}[$\hspace{10pt}$]
\item[(a)]\textbf{Find users to be scheduled in Step 2:} Find set $\textrm{DL}_{\textrm{TBS}}$ of DL users that are outside regions $A_k$ and $B_k$, for all $k$, or that are inside these regions but were not scheduled in Step 1. Find set $\textrm{UL}_{\textrm{TBS}}$ of UL users that are outside regions $A_k$ and $B_k$, for all $k$, or that are inside these regions but were not scheduled in Step 1.
\item[(b)] \textbf{Initialize variables:} Update frequency resource initial index: $f_0=f$. Initialize total number of users to be scheduled: $\textrm{TDL} = |\textrm{DL}_{\textrm{TBS}}|$ and $\textrm{TUL} = |\textrm{UL}_{\textrm{TBS}}|$.
\item[(c)]\textbf{switch: \textit{procedure}} 
\item[] \textbf{\underline{case}: \textit{FDregrand}}  \\ 
Initialize user counter $l=1$.\\
\textbf{while} $\textrm{TDL} \neq 0$ or $\textrm{TUL} \neq 0$.\\
- Update frequency resource counter:$f = f_0+l$. \\
- Assign full-duplex mode: assign frequency resource $f$ to a DL user and an UL user, $x_{\textrm{DL}_{\textrm{TBS}}(l),f} = 1$, $x_{\textrm{UL}_{\textrm{TBS}}(l),f} = 1$. \\
- Update $l=l+1$. Update $\textrm{TDL} = \textrm{TDL}-1$ and $\textrm{TUL} = \textrm{TUL}-1$.\\
\textbf{endwhile}\\
\textbf{\underline{case}: \textit{FDregHDelse}} \\
\textbf{for} $ 1 \leq l \leq \textrm{TDL}$\\
- Update frequency resource counter:$f = f_0+l$.\\
- Assign DL half-duplex mode: assign frequency resource $f$ to a DL user, $x_{\textrm{DL}_{\textrm{TBS}}(l),f} = 1$.\\
\textbf{endfor} \\
\textbf{for} $ 1 \leq l \leq \textrm{TUL}$\\
- Update frequency resource counter:$f = f_0+\textrm{TDL}+l$.\\
- Assign UL half-duplex mode: assign frequency resource $f$ to an UL user, $x_{\textrm{UL}_{\textrm{TBS}}(l),f} = 1$.\\
\textbf{endfor} 

\end{enumerate}
\end{algorithm}

\section{Performance Evaluation}
Let us now study interference and spectral efficiency in a simple scenario based on real and simulated measurements.

\subsection {System Model and Parametrization}
We assume a single BS, which can operate in half-duplex or full-duplex mode. In the latter case, we assume perfect self-interference cancellation at the BS. Users are randomly distributed over the complete radio map and their UEs operate in half-duplex, each using a single antenna.

To create the database for spatial isolation according to Section~\ref{subsec:DatabaseCreation}, we use real-world measurements from \cite{MOM}. In particular, we focus on the path loss towards the strongest serving BS and isolate a specific area in this radio map for clear representation. This area is shown in Fig. \ref{fig:RadioMap} and further parameters of the radio map are given in Table~\ref{tab:SimParam}. 

\begin{figure}
    \centering
    \begin{subfigure}{0.22\textwidth}
        \centering
        \includegraphics[height=1.5in]{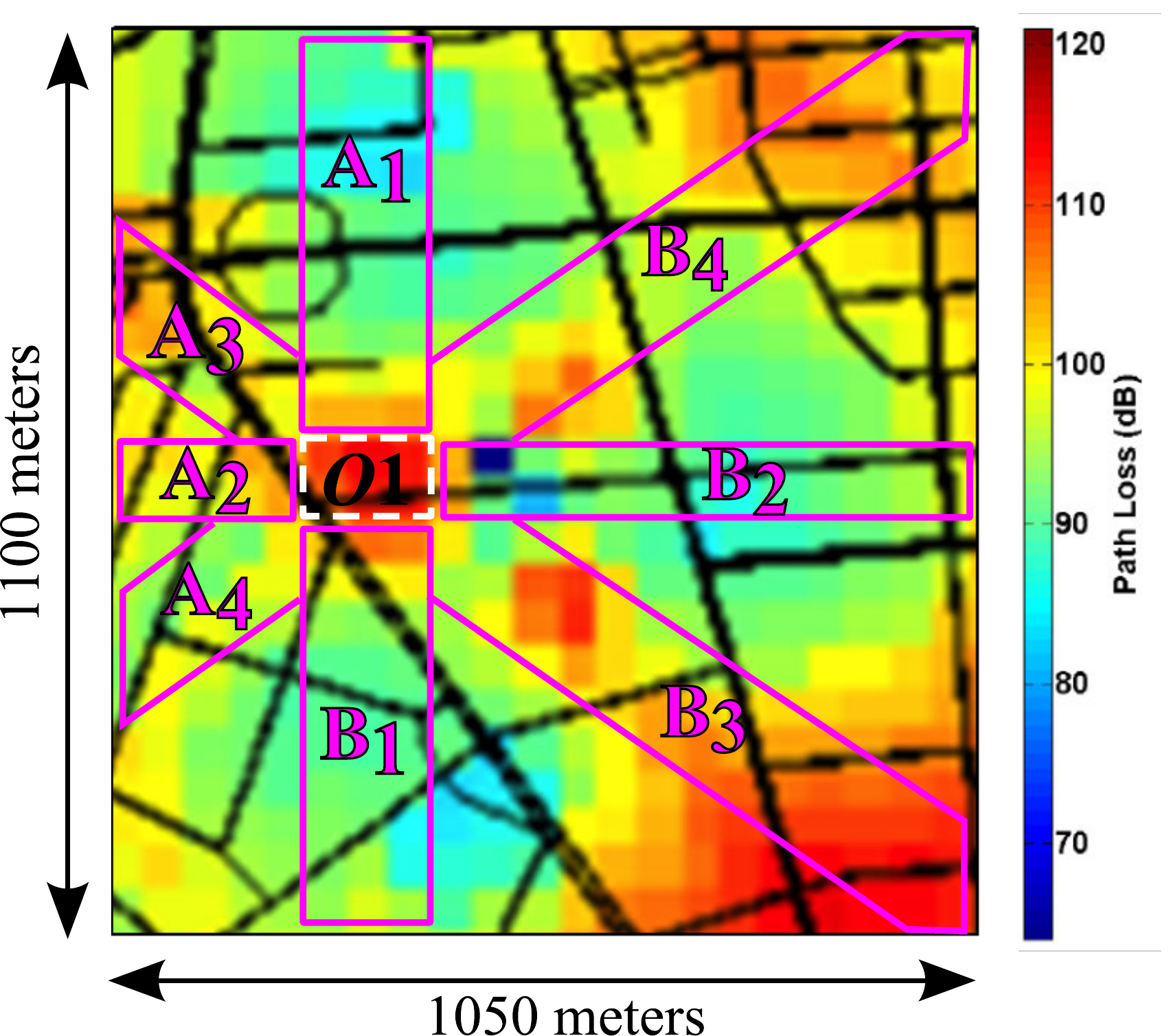}
        \caption{Four pairs of regions around obstruction $O1$.}
        \label{fig:radiomapO1}
    \end{subfigure} \quad
    \begin{subfigure}{0.22\textwidth}
        \centering
        \includegraphics[height=1.5in]{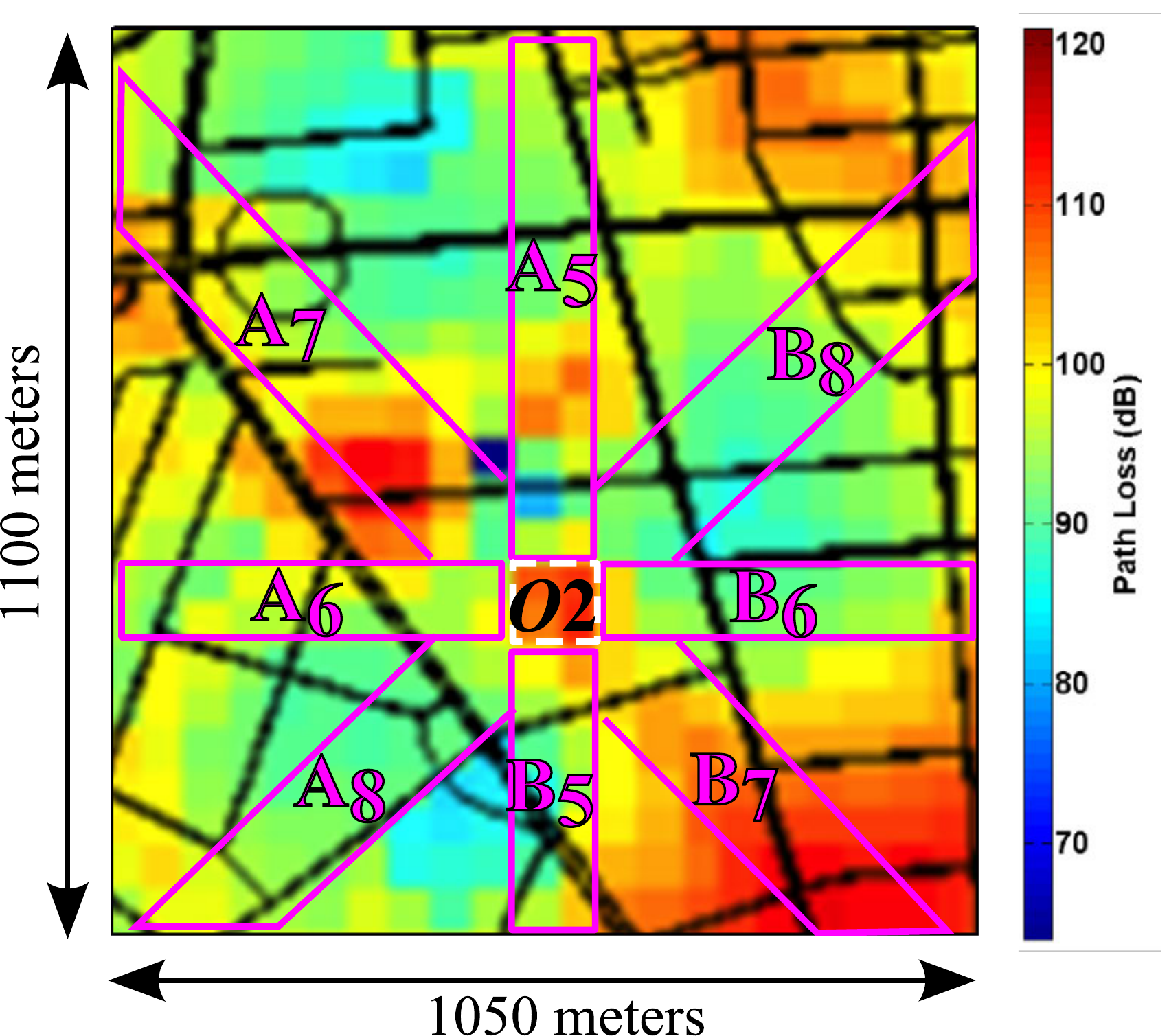}
        \caption{Four pairs of regions around obstruction $O2$.}
        \label{fig:radiomapO2}
    \end{subfigure}\qquad
    \caption{Eight pairs of regions ($A_k$, $B_k$), $k=1,2, ... 8$ around two obstructions (buildings).}
    \label{fig:RadioMapSims}
\end{figure}

In this area, we identify the two obstructions $O1$ and $O2$ resulting in the spatially isolated regions in Fig.~\ref{fig:radiomapO1} and Fig.~\ref{fig:radiomapO2}. Over both obstructions, 8 pairs of spatial isolated regions are identified where any of these pairs is fully specified by $(A_k,B_k)$ with the set of boundary coordinates $A$ and $B$ and index $k=1,\dots,8$.


Path loss between BS and UEs is available from the radio map but path loss between two arbitrary UEs $i$ and $j$ is not included. We, thus, compute this path loss, $p_{i,j}$, from 3GPP-compliant path loss models \cite{3GPP1,3GPP2} that depend on the inter-user distance $d_{i,j}$ and carrier frequency $f_{i,j}$ as shown in Table~\ref{tab:SimParam}. For the case where users $i$ and $j$ belong to an identified pair $k$ of spatially isolated regions (UE ${i}\in A_k$ and UE $j \in B_k$) then we calculate the inter-user path loss as $\max \left\{ \alpha_{k},p_{i,j} \right\}$. The constant $\alpha_{k}$ reflects the attenuation between the $k$-th pair of spatially isolated regions and is estimated to be $\alpha_{k}=140$~dB for all $k$. We justify this choice by the large size of obstructions $O1$ and $O2$ (each at least 100 m per edge) and the considerable number of walls within these buildings. This leads to a high attenuation of traversing radio signals.


Table~\ref{tab:SimParam} summarizes the main simulation parameters and propagation models.
\begin{table}[h]
			\begin{center}
				\caption{Main simulation parameters}
				\label{tab:SimParam}
					\begin{tabular}{|c|c|}
						\hline
						\textbf{Parameter }												& \textbf{Value}\\
						\hline\hline
						Frequency band $f_1$							& $2100- 2180$ MHz \\
						\hline
						Frequency band $f_2$							& $1900 - 1980$ MHz \\
						\hline
						$f_1$ and $f_2$ bandwidth					& $80$ Mhz \\
						\hline 
						Path Loss BS <-> UE 								& from radio map \cite{MOM} \\
						\hline 
						Path Loss UE$i$ <-> UE$j$ 	if d < 50 m		&  $p_{i,j} (dB) = 20  \log _{10}(4 \pi d_{i,j} f_{i,j} / c)$ \\
						& $d$ in km and $f$ in MHz \\
						& $c$ is speed of light in m/s\\
						\hline 
						Path Loss UE$i$ <-> UE$j$ 	if d > 50 m		&  $p_{i,j} (dB) =38.32 \log _{10}(d_{i,j})$\\
						& \hspace{.75cm}$+21 \log _{10}(f_{i,j})+61.6$  \\
						& $d$ in km and $f$ in MHz\\
						\hline 
						Number of DL UE & 200 \\
						\hline 
						Number of UL UE & 200 \\
						\hline 
						UE placement & Uniformly random \\
						\hline 
						BS Tx power & $46$ dBm and $20$ dBm\\
						\hline 
						Antenna height from mean rooftop & 10 m\\
						\hline 
						Noise floor & $-100$ dBm \\
						\hline 
						UE Tx power & $20$ dBm\\
						\hline
						Number of isolated regions & 8 pairs\\
						\hline 
						$\alpha$ & 140 dBm\\
						\hline 
						Radio map area size & $1050 \times 1100$ m\\
						\hline
						Radio map pixel size & $50 \times 50$ m\\
						\hline
						Radio map data source & Measurements and ray tracing \cite{MOM}\\
						\hline
					\end{tabular}
			\end{center}
		\end{table}	

\subsection{Simulated Schemes}
We compare four different schemes in order to study the performance of our proposed inter-user interference coordination method.

\begin{figure}[t]
    \centering
    \begin{subfigure}[t]{0.15\textwidth}
        \centering
        \includegraphics[width=1in]{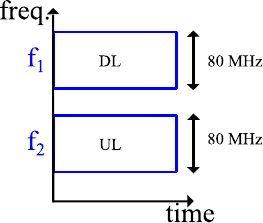}
        \caption{Half-duplex.}
        \label{fig:FrequencyHD}
    \end{subfigure}~~~
    \begin{subfigure}[t]{0.15\textwidth}
        \centering
        \includegraphics[width=1in]{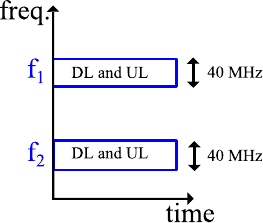}
        \caption{Full-duplex.}
        \label{fig:FrequencyFD}
    \end{subfigure}~~~
    \begin{subfigure}[t]{0.15\textwidth}
        \centering
        \includegraphics[width=1in]{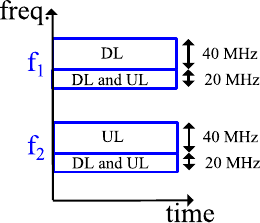}
        \caption{Hybrid-duplex.}
        \label{fig:FrequencyHybrid}
    \end{subfigure}
    \caption{Frequency utilization in simulated schemes.}
    \label{fig:Freq}
\end{figure}

\textbf{1- Baseline, half-duplex operation (HD):} In this configuration, the BS performs conventional Frequency Division Duplex (FDD) operation with frequency band $f_1$ used for uplink (UL) and frequency band $f_2$ for downlink (DL) as shown in Fig.~\ref{fig:FrequencyHD}. Inter-user interference does not occur since the users are assigned to orthogonal frequency resources. Equally for the 200 DL and 200 UL users, we assume 400~kHz of bandwidth per user, leading to 80~MHz for DL and UL, respectively and to 160 MHz in total. Frequencies are assigned randomly.\\
\textbf{2- Baseline, full-duplex operation (FDrand):} The BS performs full-duplex (FD) mode in all available frequencies with a random assignment for attached UL and DL users. Since the UL and DL users share same frequencies, the total bandwidth used is 80 MHz as shown in Fig.~\ref{fig:FrequencyFD}.\\
\textbf{3- Proposed full-duplex operation (FDregrand):} The BS uses the \textit{FDregrand} assignment heuristic explained in Section~\ref{subsec:heurstics}. The total bandwidth used is 80 MHz as shown in Fig.~\ref{fig:FrequencyFD}. Notice that this total bandwidth is equal to the \textit{FDrand} scheme. The only difference between the two schemes is the way the frequencies are assigned.\\
\textbf{4- Proposed hybrid duplex operation (FDregHDelse):} The BS uses the \textit{FDregHDelse} assignment heuristic explained in Section~\ref{subsec:heurstics}. With such a scheme, the total bandwidth used will be greater than 80 MHz but less than 160 MHz. A possible realization of 120 MHz bandwidth is illustrated in Fig.~\ref{fig:FrequencyHybrid}.

\subsection{Simulation Results}

\subsubsection {Inter-User Interference Mitigation}
\begin{figure}
	\begin{center}
	\includegraphics[width=3.1in]{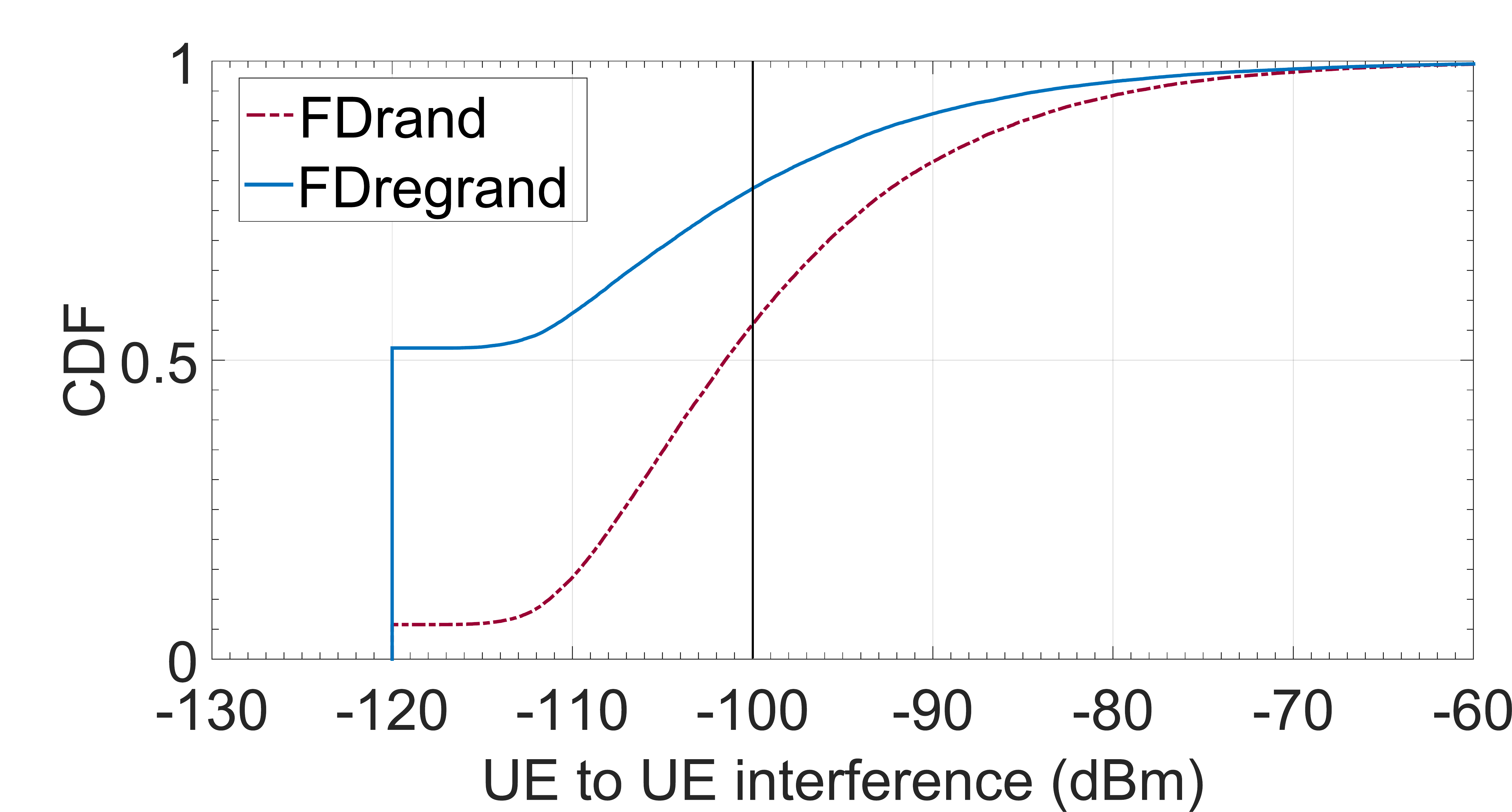}
	\caption{CDF of inter user interference.}
	\label{fig:CDFinterference}
	\end{center}
\end{figure}

We first focus on analyzing how our proposed method reduces the inter-user interference. Recall that the noise floor of the system is set at -100 dBm. Hence, when the inter-user interference increases beyond -100~dBm, then the interference experienced by the DL user becomes larger than the noise floor and the performance degradation is dominated by interference. From Fig.~\ref{fig:CDFinterference}, we observe that for the \textit{FDrand} scheme, the probability that UE-UE interference is larger than -100 dBm is 44\%. For the \textit{FDregrand} scheme, this probability is 22\%. Hence, our proposed method substantially reduces the number of links that are dominated by inter-user interference. Specifically, using the regions with the \textit{FDregrand} scheme mitigates the inter-user interference such that the probability of interference dominating over the noise floor decreases by 50\% compared to the reference \textit{FDrand }scheme. Note that this decrease will be even more significant for lower noise floor.\\
In addition, we notice a jump of the \textit{FDregrand} CDF at $-120$~dB. Recall that in our simulations the value of $\alpha=140$~dB and the UEs transmit at 20 dBm. Hence, for FD links with users in the identified regions, the inter-user interference will be at most $-120$~dBm. Many users in the regions experience this value of interference since the path loss between users in the regions is $\max\{140, p_{i,j}\}$~dB. Hence, the fact that several users experience this level of interference results in this jump. The fact that the jump at $-120$~dBm is almost not present in the \textit{FDrand} results shows that a random assignment is not capable of harnessing the spatial isolation provided by the geographical context. Notice that the \textit{FDregrand} scheme took advantage of the spatially isolated regions such that there is approximately a 50\% probability that the inter-user interference is less than or equal to $-120$~dBm. In contrast, for the \textit{FDrand} scheme, this probability is only about 5\%.

\subsubsection{Spectral Efficiency}

\begin{figure}[t]
    \centering
    \begin{subfigure}[t]{0.45\textwidth}
        \centering
        \includegraphics[width=2.8in]{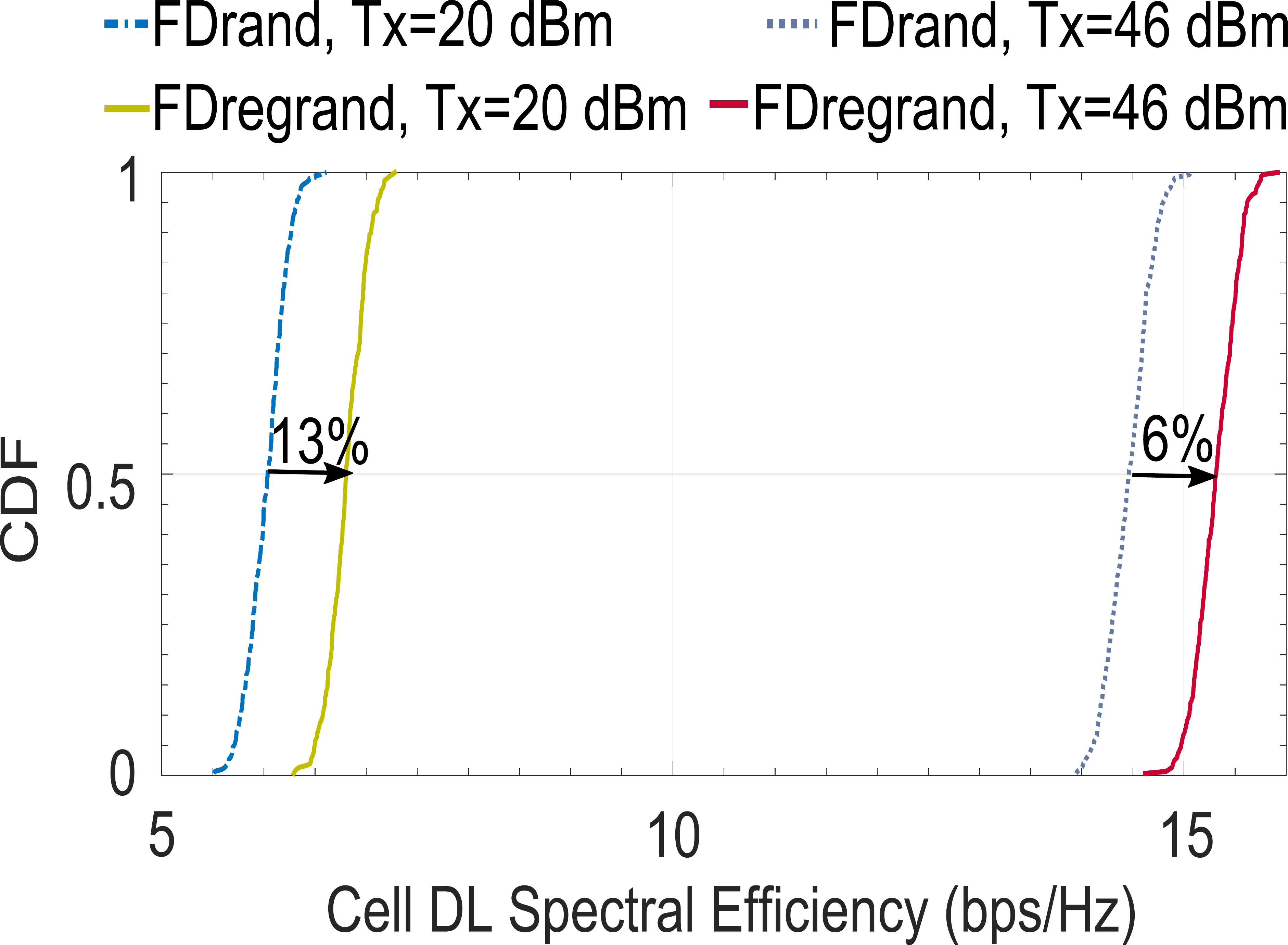}
        \caption{CDF of downlink spectral efficiency for the entire cell.}
        \label{fig:CDFSpecEffDLCellAll}
    \end{subfigure}\\
    \centering    
    \begin{subfigure}[t]{0.45\textwidth}
        \centering
        \includegraphics[width=2.8in]{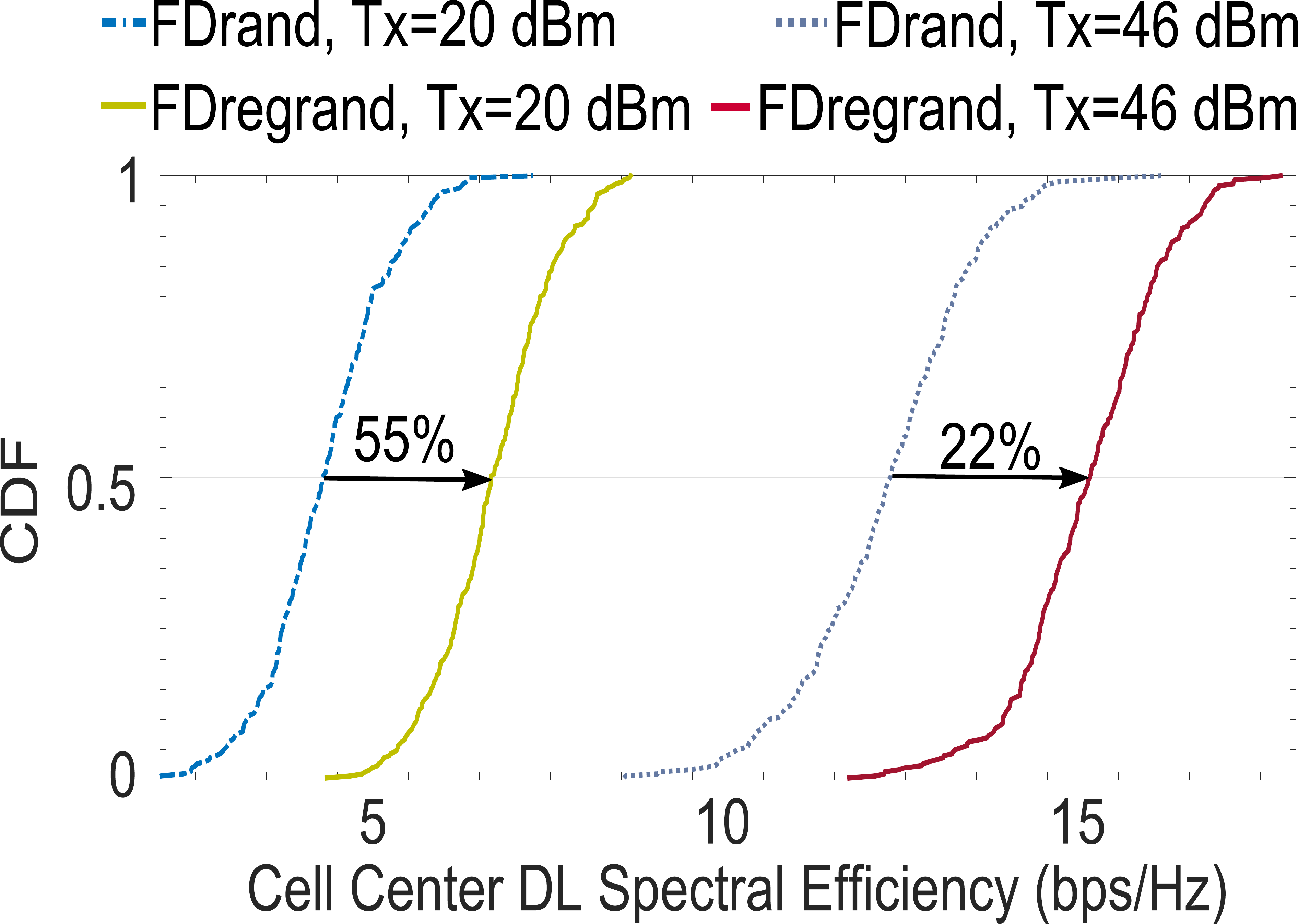}
        \caption{CDF of downlink spectral efficiency for cell center.}
        \label{fig:CDFSpecEffDLCellCenter}
    \end{subfigure}
    \caption{Simulation results comparing \textit{FDrand} and \textit{FDregrand} schemes.} 
    \label{fig:CDFSpecEffDL_FD}
\end{figure}
\begin{figure}[t]
    \centering
    \begin{subfigure}[t]{0.45\textwidth}
        \centering
        \includegraphics[width=2.8in]{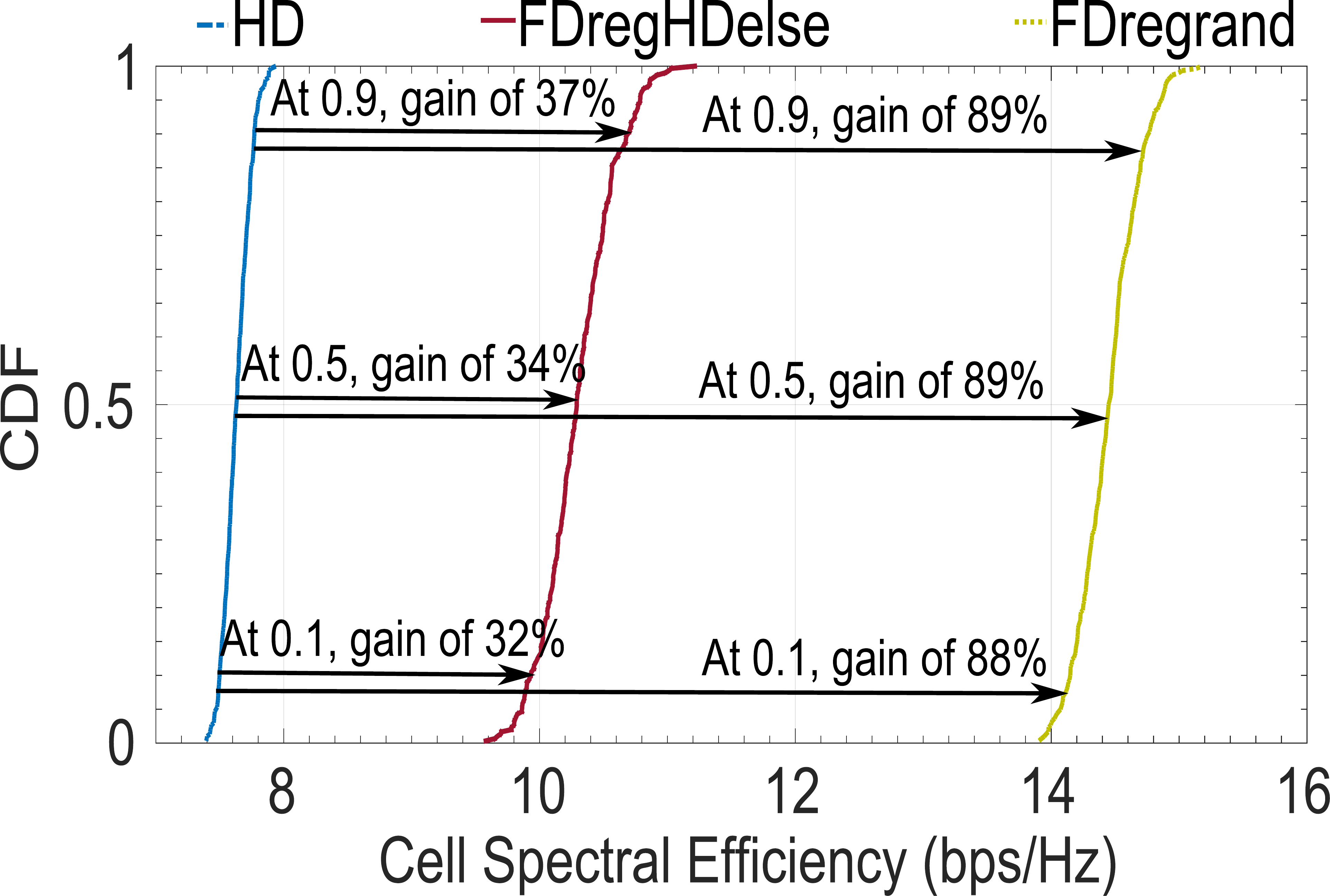}
        \caption{CDF of spectral efficiency for the entire cell when the base station has transmission power of 20~dBm.}
        \label{fig:CDFSpecEffULDCellAllS20dBm}
    \end{subfigure}\\
    \begin{subfigure}[t]{0.45\textwidth}
        \centering
        \includegraphics[width=2.8in]{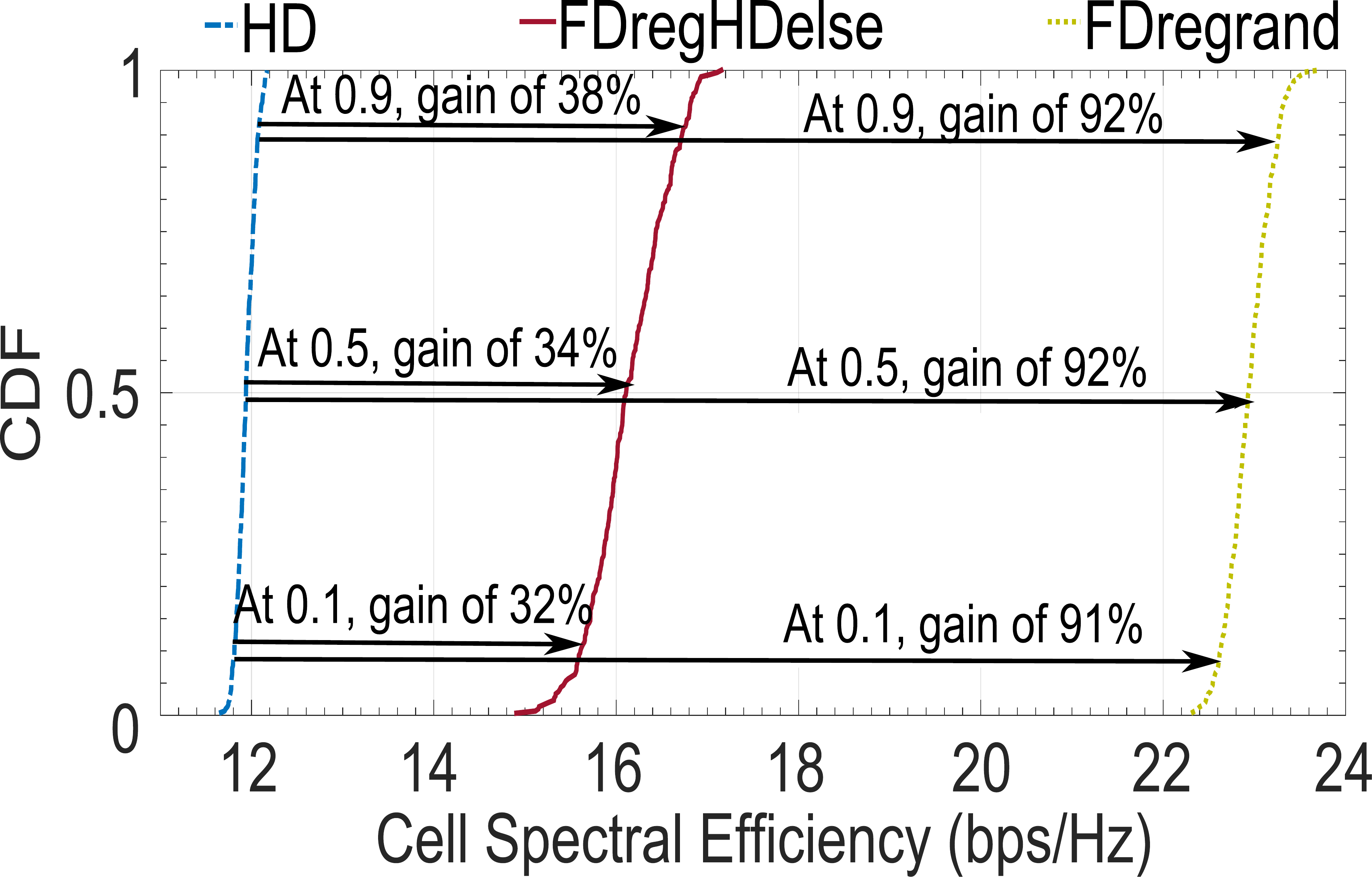}
        \caption{CDF of spectral efficiency for the entire cell when the base station has transmission power of 46~dBm.}
        \label{fig:CDFSpecEffULDLCellAllBS46dBm}
    \end{subfigure}
    \caption{Simulation results comparing half-duplex with \textit{FDregrand} and \textit{FDregHDelse}.} 
    \label{fig:CDFSpecEffDL_HDFD}
\end{figure}

In Fig.~\ref{fig:CDFSpecEffDLCellAll} and Fig.~\ref{fig:CDFSpecEffDLCellCenter} we compare the spectral efficiency of the \textit{FDrand} and \textit{FDregrand} schemes. Both schemes have the same UL performance so we focus our comparison on the DL. We consider BS transmitted power of $20$~dBm (Pico cell) and $46$~dBm (Macro cell). The results in Fig.~\ref{fig:CDFSpecEffDLCellAll} are computed for all the DL users in the cell while the results in Fig.~\ref{fig:CDFSpecEffDLCellCenter} reflect only the users in the cell center. Here, only DL users and co-channel UL users within a 300 m radius from the BS are considered.\fnsv{Something is wrong with the captions of the sub-figures in Fig. 6: There should be two plots for the cell center but the captions indicate only one. I think one of the plots was erroneously tagged to be for the complete cell.}

In Fig.~\ref{fig:CDFSpecEffDLCellAll} and Fig.~\ref{fig:CDFSpecEffDLCellCenter}, we observe that the \textit{FDregrand} has higher spectral efficiency than the \textit{FDrand} scheme. The gain obtained across the entire cell is modest. From Fig.~\ref{fig:CDFSpecEffDLCellAll}, we compute this gain to be 6\% for a BS transmit power of 46~dBm and 13\% for a BS transmit power of 20~dBm. In contrast, the gain is more significant if we observe only cell center links. From Fig.~\ref{fig:CDFSpecEffDLCellCenter}, we compute that the gain is 22\% for a BS transmit power of 46~dBm and 55\% for a BS transmit power of 20~dBm. 

This significant increase in gain for cell center users results from the fact that the closer users are to the BS, the more full-duplex communication suffers from inter-user interference. Also, the closer UL and DL users are to the base station, the more likely it is that they are close to each other and, thus, generating high inter-user interference power among themselves. The \textit{FDregrand} scheme uses the spatially isolated regions to carefully choose which UL-DL users can share the same frequency band and this results in an effective decrease of the inter-user interference experienced at the cell center. Here, enabling full-duplex is of particular interest due to the small average proximity of the users to the BS, which is a favorable condition for full-duplex operation since it can simplify the implementation of self-interference cancellation~\cite{SAB14}. However, cell-center full-duplex operation also  results small average proximity between users and high loss in spectral efficiency from inter-user interference. The results in Fig.~\ref{fig:CDFSpecEffDLCellCenter} show that our proposed scheme can effectively reduce this interference in order to achieve 55\% improvement over a scheme that does not exploit the spatially isolated regions.

\subsubsection{Comparison with Half-Duplex}

The results in Figs.~\ref{fig:CDFSpecEffULDCellAllS20dBm} and \ref{fig:CDFSpecEffULDLCellAllBS46dBm} compare the performance of half-duplex against the performance of the hybrid-duplex scheme that we labeled \textit{FDregHDelse} and against the performance of the \textit{FDregrand} scheme. We observe that the use of full-duplex and the spatially isolated regions improves the spectral efficiency over the performance of \textit{HD}. The results in Fig.~\ref{fig:CDFSpecEffULDCellAllS20dBm} are for a BS transmit power of 20~dBm and the results in Fig.~\ref{fig:CDFSpecEffULDLCellAllBS46dBm} are for a BS transmit power of 46~dBm. An observation of the medians shows that for both transmission powers, the hybrid scheme achieves 34\% gain over HD. Also, for both cases, the \textit{FDregrand} scheme further improves to achieve between 89\% and 92\% gain over HD.

\section{Conclusions}\label{sec:concl}
We described in this paper a new inter-user interference coordination scheme in cellular networks with full-duplex. Our scheme assigns simultaneous uplink and downlink transmissions to users in areas that are separated by large obstacles. The attenuation from such obstacles, isolates the simultaneous transmissions from each other and, thus, allows to minimize interference. The proposed assignment scheme adds only low computational complexity and signaling overhead to the Radio Access Network (RAN), while segmenting radio maps into spatial isolated areas can be performed offline, in arbitrary data centers. 


The initial performance results of this approach are impressive. For small and large cells, outstanding spectral efficiency gains can be shown that are clearly due to the reduction of inter-user interference. Especially in the cell center -- where inter-user interference dominates over path loss -- very high increases in spectral efficiency are provided.

We conclude that the described inter-user interference coordination scheme is a promising example of exploiting geographical context information in RAN. The outstanding gains, the low computational complexity, and the increasing availability of radio maps and user positions, make this approach a promising candidate to support full-duplex in future RAN generations.


\def\bibfont{\footnotesize}
\bibliographystyle{./IEEEtran}
\bibliography{biblioFD}

\end{document}